\documentstyle[preprint,prc,aps,floats,epsf]{revtex}
\def\thalf{{\textstyle{\frac{1}{2}}}}
\def\tquar{{\textstyle{\frac{1}{4}}}}
\def\thrhalf{{\textstyle{\frac{3}{2}}}}

\begin{document}


\preprint{NUC-MINN-96/17-T}

\title{Pion--Nucleon Scattering at Low Energies}

\author{Paul J. Ellis and Hua-Bin Tang}
\address{School of Physics and Astronomy,
     University of Minnesota, Minneapolis, MN\ \ 55455.}

\date{\today}
\maketitle

\begin{abstract}
We study pion--nucleon scattering  at tree level with a chiral lagrangian 
of pions, nucleons, and $\Delta$-isobars using a $K$-matrix unitarization
procedure. Evaluating the scattering amplitude to order $Q^2$,
where $Q$ is a generic small momentum scale, 
we obtain a good fit to the experimental phase shifts  
for pion center-of-mass kinetic energies up to $50\,$MeV. 
The fit can be extended to 150 MeV when we include  the
order-$Q^3$ contributions.
Our results are independent of the off-shell $\Delta$ parameter.
\end{abstract}

PACS number(s): 11.80.-m, 12.39.Fe, 13.75.Gx

\vspace{20pt}
%


Pion--nucleon ($\pi N$) scattering is a fundamental hadronic process
for which a large amount of data is available and it is important to
understand this as completely as possible.
Several relativistic models 
\cite{OLSSON75,BOFINGER91,PEARCE91,GROSS93,GOUDSMIT94} 
exist which provide reasonably good fits to the experimental 
phase shifts. These models consider the $N$, $\pi$ and 
$\Delta$-resonance fields, the isoscalar-scalar $\phi$ (in some cases
implictly via a power series expansion),
the $\rho$ meson and sometimes higher resonances, 
although these play a minor role. Our interest here is to 
examine whether a model which contains the minimal 
number of fields, namely the $N$, $\pi$ and $\Delta$ ,
can yield equally good fits. Thus we effectively integrate out any other 
fields. For example, provided the center-of-mass (c.m.) energy is not too 
high, we can expand the $\rho$ propagator as 
$(m_\rho^2-t)^{-1}=m_\rho^{-2}(1+t/m_\rho^2+\cdots)$, where the Mandelstam 
variable $t=(q-q')^2$ and $q$ and $q'$ are the initial and final pion 
c.m. four-momenta. The series of terms can be absorbed into contact 
interactions in the 
lagrangian and it is clearly important to employ
the most general set of such contact interactions which is 
consistent with the symmetries of quantum chromodynamics.

While the $\Delta$ degree of freedom plays an important role in $\pi N$ 
scattering, the $Z$ parameter that specifies the form of the 
$\pi N\Delta$ vertex has been controversial, see the discussion 
of Benmerrouche {\it et al.} \cite{BEN89}. Most of the papers cited above
fit the $Z$ parameter to the $\pi N$ data. This is unsatisfactory since, 
as we showed recently \cite{TE96}, the scattering is independent of $Z$ 
if the lagrangian contains the most general set of contact terms (we 
demonstrate this explicitly below). Thus results which depend on $Z$
indicate that the contact terms have been implicitly 
constrained, whereas it is clearly preferable to employ 
a general lagrangian and allow the data itself to impose 
constraints.

We would like to employ a lagrangian which explicitly embodies chiral 
symmetry since this is known to be a fundamental symmetry at low energies.
Such an approach was first taken by Peccei \cite{PECCEI63} to calculate the 
scattering lengths and this paper represents a modern extension of his work
to study the phase shift data. In order to systematically enumerate the 
lagrangian we can be guided by Weinberg's power counting arguments
\cite{WEINBERG90}. For this purpose we identify a generic small-momentum 
scale $Q$. This is of the order of the pion three-momentum or the pion mass
and therefore much smaller than the scale of the nucleon or the delta mass.
Then according to
the power counting, a Feynman tree diagram without loops
contributes to $\pi N$ scattering at order $Q^\nu$ with
\begin{equation}
      \nu = 1 + \sum_{i} V_i \Big (d_i + \thalf n_i -2\Big) 
                 \ , \label{eq:chcnt}
\end{equation}
where $V_i$ is the number of
vertices of type $i$ characterized by $n_i$ baryon
fields and $d_i$ pion derivatives or $m_\pi$ factors. 
This suggests that we associate $d_i + \thalf n_i$
powers of $Q$ to a term of type $i$ in the lagrangian \cite{FSTnew}.
Also, Krause\cite{KRAUSE90} argues that $i\rlap/{\mkern-2mu {\cal D}} -M$
is of $O(Q)$, as is a single factor of $\gamma_5$
(note $\gamma_\mu\gamma_5$ is of $O(1)$). 
Although we naively count $\gamma_5$ as $O(Q)$ for organizing
the lagrangian, we shall show later that this counting is not
precise.  Chiral symmetry ($SU(2)\otimes SU(2)$), 
Lorentz invariance, and parity constrain
the possible $\pi N$ interactions and these can be found in
Ref.~\cite{GASSER88}. For interactions involving the $\Delta$ 
isobar we use the notation of our
previous paper\cite{TE96} and follow the discussion therein.
We write the lagrangian up to quartic order as
the sum of order $Q^2$, $Q^3$, and $Q^4$ parts:
%
${\cal L} = {\cal L}_2 + {\cal L}_3 + {\cal L}_4 \ .$
%

The order $Q^2$ part of the lagrangian is
\begin{eqnarray}
  {\cal L}_2 &=& \overline{N} ( i\rlap/{\mkern-2mu {\cal D}}
               +g_{\rm A}\gamma^\mu \gamma_5 a_\mu - M ) N 
           +\tquar f_\pi^2  {\rm tr}\, (\partial_\mu U^\dagger
                         \partial^\mu U)
           +\tquar m_\pi^2f_\pi^2 {\rm tr}\,(U + U^\dagger -2)
       \nonumber  \\
      & & 
          +  \overline{\Delta}_\mu^a
           \Lambda^{\mu\nu}_{ab} \Delta_\nu^b
       + h_{\rm A} \Big ( \overline{\bbox{\Delta}}_\mu
         \bbox{\cdot a}_\nu \Theta^{\mu\nu} N
       + \overline{N}
           \Theta^{\mu\nu} \bbox{a}_\mu\bbox{\cdot \Delta}_\nu \Big)
        + \tilde{h}_{\rm A}
            \overline{\Delta}_\mu^{\, a}
                  \rlap/{\mkern-1mu a }\gamma_5 
                    \Delta^\mu_a   \ .   \label{eq:L2}
\end{eqnarray}
where the pion field arises in
$U(x) =\exp (2i\pi(x)/ f_{\pi})$
with $\pi \equiv \thalf\bbox{\pi\cdot \tau}$ and the axial vector field
$a_{\mu} =\partial_{\mu}\pi/f_\pi+\cdots\,$, while the vector field
$v_{\mu} = -\thalf i[\pi,\partial_{\mu}\pi]/f^2_\pi+\cdots\,$.
The trace is taken over the isospin matrices and the 
covariant derivative on the nucleon field  is
${\cal D}_\mu N = \partial_\mu N + i v_\mu N$.
As regards the $\Delta$, the kernel tensor in the  
kinetic-energy term is
\begin{equation}
    \Lambda^{\mu\nu} = -(i \rlap/{\mkern-2mu {\cal D}} 
                - M_\Delta)g^{\mu\nu} 
              +i(\gamma^\mu {\cal D}^\nu +\gamma^\nu {\cal D}^\mu)
              -\gamma^\mu(i\rlap/{\mkern-2mu {\cal D}}+M_\Delta)\gamma^\nu
                        \ .        
\end{equation}
Here we have chosen the standard parameter $A=-1$, because
it can be modified by redefinition of the $\Delta$ field with no
physical consequences \cite{NATH70}. The covariant derivative is 
\begin{equation}
{\cal D}_\mu \bbox{\Delta}_\nu
            = \partial_\mu\bbox{\Delta}_\nu
                      + i  v_\mu  \bbox{\Delta}_\nu
                      -   \bbox{v}_\mu \times \bbox{\Delta}_\nu
                            \ , \label{eq:covder}
\end{equation}
in which $\bbox{\Delta}_\mu = \bbox{T} \Delta_\mu$, with $T^a$ 
the standard $2\times 4$
isospin $\thrhalf$ to $\thalf$  transition matrices.
The off-shell $Z$ parameter appears in
%
$
  \Theta_{\mu\nu} = g_{\mu\nu} -
        \Big( Z + \thalf \Big)\gamma_\mu\gamma_\nu \,
$.
%
We have simplified the $\pi\Delta\Delta$ interaction
in Eq. (\ref{eq:L2}) by choosing the physically irrelevant parameters 
$Z_2=-\thalf$ and $Z_3=0$ (see Ref.~\cite{TE96}); this term 
does not contribute to the scattering amplitude at tree level.

The order $Q^3$ part of ${\cal L}$ is
\begin{eqnarray}
  {\cal L}_3 &=&  { \beta_\pi \over M} \overline{N} N 
         {\rm tr}\, (\partial_\mu U^\dagger \partial^\mu U)
          -{\kappa_\pi\over M} \overline{N}
               v_{\mu\nu}\sigma^{\mu\nu} N
                     \nonumber  \\
      & & +{\kappa_1\over 2 M^2} i\overline{N}
               \gamma_\mu 
               \stackrel{\leftrightarrow}{\cal D}_\nu N
             {\rm tr}\, (a^\mu a^\nu)
          +{\kappa_2\over M} m_\pi^2 \overline{N} N\, 
            {\rm tr}\,(U + U^\dagger -2) +\cdots
                   \ ,     \label{eq:L3}
\end{eqnarray}
where the dots  represent terms that do not 
contribute to the $\pi N$ scattering amplitude up to $O(Q^3)$
and we have defined
\begin{eqnarray}
   \stackrel{\leftrightarrow}{\cal D}_\mu
      &=& {\cal D}_\mu -
         (\stackrel{\leftarrow}{\partial}_\mu - iv_{\mu}) \ ,  \\
    v_{\mu\nu} &=& \partial_\mu v_\nu -\partial_\nu v_\mu 
                  + i [v_\mu, v_\nu]  \ . 
\end{eqnarray}
We have also applied naive dimensional analysis\cite{GEORGI93}
to factor out the dimensional factors so that the parameters are expected 
to be of order unity.

Finally, the order $Q^4$ part of ${\cal L}$ is
\begin{eqnarray}
  {\cal L}_4 &=& 
             {\lambda_1\over M} m_\pi^2 \overline{N} \gamma_5
                (U - U^\dagger ) N
             + {\lambda_2\over M^2} \overline{N}
                        \gamma^\mu D^\nu v_{\mu\nu} N
                     \nonumber  \\
      & & 
           +{\lambda_3\over M^2} m_\pi^2 \overline{N} \gamma_\mu
               [a^\mu, \, U - U^\dagger] N
          +{\lambda_4\over 2 M^3} i\overline{N}
               \sigma_{\rho \mu} 
               \stackrel{\leftrightarrow}{\cal D}_\nu N
             {\rm tr}\, (a^\rho D^\mu a^\nu)
                     \nonumber  \\
      & &
          +{\lambda_5\over 16M^4} i\overline{N}
         \gamma_\rho 
            \{\stackrel{\leftrightarrow}{\cal D}_\mu,
            \stackrel{\leftrightarrow}{\cal D}_\nu\} \tau^a N 
          \,{\rm tr}\,(\tau^a[D^{\rho}a^\mu,a^\nu])
            +\cdots
                   \ ,     \label{eq:L4}
\end{eqnarray}
where the braces denote an anticommutator and 
\begin{equation}
    D_\mu a_{\nu} = \partial_\mu a_{\nu}
                            + i [v_\mu, a_{\nu}] \ ,\quad
    D^\sigma v_{\mu\nu} = \partial^\sigma v_{\mu\nu}
                            + i [v^\sigma, v_{\mu\nu}] \ .
\end{equation}
Again the dots  represent  terms that do not 
contribute to the $\pi N$ scattering amplitude up to $O(Q^3)$,
such terms include the usual fourth-order pion lagrangian.

Using the pion and nucleon equations of motion 
\cite{WEINBERG90,GEORGI91,ARZT95}, 
we have simplified 
the contact terms listed in Ref.~\cite{GASSER88}.
For example, we reduce the $O(Q^3)$ term 
$ \overline{N}   \stackrel{\leftrightarrow}{\cal D}_\mu 
               \stackrel{\leftrightarrow}{\cal D}_\nu\! N
             \, {\rm tr}\, (a^\mu a^\nu)
$
to the sum of the $O(Q^3)$ $\kappa_1$ term, 
the $O(Q^4)$ $\lambda_4$ term, and higher-order terms which we omit.
As a result we have the  minimum number of independent terms 
contributing to the $\pi N$ scattering amplitude up to $O(Q^3)$. 
As we have remarked, the isoscalar-scalar $\phi$
and isovector-vector $\rho$ fields as given in Ref.~\cite{FSTnew}
have been integrated out. Their effects show up 
in the contact terms $\beta_{\pi},\ \kappa_2$ and 
$\lambda_2$. For example, in terms of the 
$\rho\pi\pi$ coupling ($g_{\rho\pi\pi}$) and the
$\rho N N$ coupling ($g_\rho$), the rho gives a contribution
to the $\lambda_2$ parameter of 
$-2 g_{\rho\pi\pi}g_\rho M^2f_\pi^2/m_\rho^4$.

\begin{figure}
 \setlength{\epsfxsize}{3in}
  \centerline{\epsffile{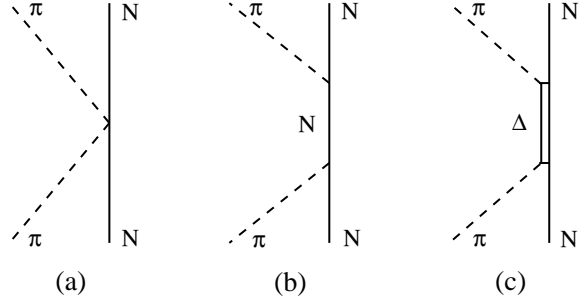}}
\vspace*{.2in} 
\caption{Tree Feynman diagrams for $\pi N$ scattering:
(a) contact terms, (b) nucleon exchange, (c) $\Delta$ exchange.
Crossed diagrams  are not shown. }
 \label{fig:one}
\vspace*{.4in}
\end{figure}

In Fig.~\ref{fig:one} we show the tree Feynman diagrams
for $\pi N$ scattering. The
crossed diagrams for Figs.~\ref{fig:one}(b) and \ref{fig:one}(c)
are suppressed. The lagrangian ${\cal L}_2$ gives
contributions to the $T$ matrix of $O(Q)$ from all three diagrams; 
note that the contact diagram is due to 
the Weinberg term $-\overline{N}\gamma^{\mu}v_{\mu}N$.
The  interactions in ${\cal L}_3$ and ${\cal L}_4$ 
(except for the $\lambda_1$ term) give further contributions to
Fig.~\ref{fig:one}(a) of order $Q^2$ and $Q^3$, respectively.
In  Fig.~\ref{fig:one}(b),
each vertex can be either a pseudovector $g_A$ vertex 
or  a symmetry-breaking $\lambda_1$ vertex. 
As mentioned earlier, the appearance of 
$\gamma_5$ renders the $\lambda_1$ vertex of higher order than
expected from the chiral counting of Eq.~(\ref{eq:chcnt}) so we 
have included it in ${\cal L}_4$.
The reason for this extra power of $Q$ results from
the following relation:
\begin{equation}
\overline{u}(p')\gamma_5
       {1\over \rlap/{\mkern-2mu p}
       +\rlap/{\mkern-1mu q} - M}\gamma_5 u(p) =
-{ \overline{u}(p')\rlap/{\mkern-1mu q} u(p) \over (p+q)^2-M^2  }  
                                  \ ,
\end{equation}
where $u(p)$ is the positive-energy free Dirac spinor.
Thus, with one $\lambda_1$ and one $g_A$ vertex, Fig.~\ref{fig:one}(b)
is of $O(Q^3)$; we include this contribution. With both vertices of 
$\lambda_1$ type the result 
is of $O(Q^4)$, whereas, associating an extra factor of $Q$ with each 
$\gamma_5$ as suggested by Ref.~\cite{KRAUSE90} 
and Eq. (\ref{eq:L4}), we would expect $O(Q^5)$.

We follow the 
standard notation of H\"{o}hler\cite{HOHLER83} and Ericson and
Weise\cite{EW88} to write  the $T$ matrix as
\begin{equation}
  T_{ba} \equiv \langle \pi_b | T | \pi_a\rangle
         =T^+ \delta_{ab} + \case{1}{2}[\tau_b, \tau_a] T^- \ ,
\end{equation}
where the isospin symmetric and antisymmetric amplitudes are 
\begin{equation}
    T^{\pm} = A^{\pm} + \case{1}{2}(\rlap/{\mkern-1mu q}+
                           \rlap/{\mkern-1mu q'}) B^\pm \ .
\end{equation}
Here $A^\pm$ and $B^\pm$ are functions of the Mandelstam 
invariant variables $s=(p+q)^2$, $t$, and $u=(p-q')^2$, where
$p$ is the initial nucleon c.m. momentum. They are given by 
the sum of the contributions from the contact terms in
Fig.~\ref{fig:one}(a), the nucleon exchange in Fig.~\ref{fig:one}(b),
and the $\Delta$ exchange in Fig.~\ref{fig:one}(c).
The amplitudes arising from the contact terms are
\begin{eqnarray}
  A_{\rm C}^+ & = & {2\over M f_\pi^2}
    \Big[ \beta_\pi\Big(2m_\pi^2 -t\Big) 
                 -2\kappa_2 m_\pi^2 
                 +\lambda_4 \nu^2 \Big]  \ ,
                          \\
  B_{\rm C}^+ & = & {1\over M f_\pi^2}(\kappa_1 - 2\lambda_4)\nu \ ,
                          \\
  A_{\rm C}^- & = & -{2 \kappa_\pi \over  f_\pi^2} \nu
                        \ ,         \\
  B_{\rm C}^- & = & {1\over 2 f_\pi^2}(1 + 4\kappa_\pi ) 
                   - { 1 \over  M^2 f_\pi^2}
                       \Big(\case{1}{2}\lambda_2 t 
                             +4 \lambda_3 m_\pi^2 -\lambda_5 \nu^2 
                        \Big) \ ,
\end{eqnarray}
where $\nu = (s-u)/4M$. The contributions from
the nucleon-  and $\Delta$-exchange  are well-known
(see Ref.~\cite{HOHLER83} for example). We list these 
contributions in the following for completeness.
The amplitudes arising from nucleon exchange  are
\begin{eqnarray}
  A_{\rm N}^+ & = & { M\over f_\pi^2} g_{\rm A}
           \Big ( g_{\rm A} - 4\lambda_1 {m_\pi^2 \over M^2} \Big ) \ ,
                          \\
  B_{\rm N}^+ & = & { M\over f_\pi^2} g_{\rm A}
           \Big ( g_{\rm A} - 4\lambda_1 {m_\pi^2 \over M^2} \Big )
                {\nu \over \nu_{\rm B}^2 -\nu^2} \ ,
                          \\
  A_{\rm N}^- & = & 0  \ ,
                          \\
  B_{\rm N}^- & = & -  {g_{\rm A}^2\over 2 f_\pi^2}
                 + { M\over f_\pi^2} g_{\rm A}
           \Big ( g_{\rm A} - 4\lambda_1 {m_\pi^2 \over M^2} \Big )
                {\nu_{\rm B} \over \nu_{\rm B}^2 -\nu^2} \ ,
\end{eqnarray}
where $\nu_{\rm B} = (t-2m_\pi^2)/4M$.
The amplitudes arising from $\Delta$ exchange are
\begin{eqnarray}
  A_{\Delta}^+ & = &{2 h_{\rm A}^2 \over 9 M f_\pi^2}
                      \Big [\alpha_1+ \case{3}{2} (M_\Delta+M) t \Big] 
                 {\nu_\Delta  \over \nu_{\Delta}^2 -\nu^2}
                  \nonumber \\
            & &
                 - {4 h_{\rm A}^2 \over 9 M_\Delta f_\pi^2}
             \Big [
               (E_\Delta + M)(2M_\Delta - M) +
                    \Big(2 + {M\over 2 M_\Delta}\Big) m_\pi^2
               -(2m_\pi^2 - t) Y
             \Big] 
             \ ,      \label{AplusD}       \\
  B_{\Delta}^+ & = & {2h_{\rm A}^2 \over 9 M f_\pi^2}
                      \Big[2(E_\Delta+M)(E_\Delta - 2M) +
                          \case{3}{2} t\Big]
                 {\nu \over \nu_{\Delta}^2 -\nu^2}
                 - {16 h_{\rm A}^2 \over 9 f_\pi^2}
                    {M \over M_\Delta ^2} Z^2 \nu
             \ ,            \\
  A_{\Delta}^- & = & -{h_{\rm A}^2 \over 9 M f_\pi^2}
                      \Big [\alpha_1+ \case{3}{2} (M_\Delta+M) t \Big]
                 {\nu \over \nu_{\Delta}^2 -\nu^2}
                 - {8 M h_{\rm A}^2 \over 9 M_\Delta f_\pi^2} Y \nu
             \ ,             \\
  B_{\Delta}^- & = & - {h_{\rm A}^2 \over 9 M f_\pi^2}
                      \Big[2(E_\Delta+M)(E_\Delta - 2M) +
                          \case{3}{2} t\Big]
                 {\nu_\Delta \over \nu_{\Delta}^2 -\nu^2}
                        \nonumber \\
            & &
                 +{ h_{\rm A}^2 \over 9 f_\pi^2}
                  \bigg\{ \Big(1 + {M\over M_\Delta}\Big)^2
                         + {8M\over M_\Delta} Y
                         + {2\over M_\Delta^2}
                           \Big[(2m_\pi^2 -t)Z^2 - 2m_\pi^2 Z\Big ]\bigg\}
                         \ ,  \label{BminusD} 
\end{eqnarray}
where $\nu_\Delta = (2M_\Delta^2 - s -u)/4M$, 
      $E_\Delta = (M_\Delta^2 +M^2 -m_\pi^2)/2M_\Delta$, and
\begin{eqnarray}
        \alpha_1 &=& 2(E_\Delta+M)[ M_\Delta (2E_\Delta-M)
                  +M(E_\Delta - 2M)] \ ,  \\
        Y(Z)  &=& \bigg(2 + {M\over M_\Delta} \bigg)  Z^2 
              + \bigg(1 + {M\over M_\Delta} \bigg) Z    \ .
\end{eqnarray}

Notice that in agreement with Ref.~\cite{TE96} only the nonpole terms
in the $\Delta$-exchange diagram involve the off-shell parameter $Z$. 
Therefore these contributions can be absorbed into the parameters of the 
contact terms according to
\begin{eqnarray}
 \beta_\pi(Z) &=& \beta_\pi(-\case{1}{2}) -
                {h_{\rm A}^2 \over 18} 
                \bigg[ 4Y(Z) +{M\over M_\Delta}\bigg] {M\over M_\Delta}
                           \ , \label{eq:zbeta} \\[3pt]
 \kappa_\pi(Z) &=&\kappa_\pi(-\case{1}{2}) -
                {h_{\rm A}^2 \over 9} 
                \bigg[ 4Y(Z) +{M\over M_\Delta}\bigg] {M\over M_\Delta}
                              \ , \\[3pt]
 \kappa_1(Z) &=&\kappa_1(-\case{1}{2}) +
                {4 h_{\rm A}^2 \over 9} 
                \Big( 4Z^2 -1\Big) {M^2\over M_\Delta^2}  \ , \\[3pt]
 \lambda_2(Z)  &=&\lambda_2(-\case{1}{2}) -
                {h_{\rm A}^2 \over 9} 
                \Big( 4Z^2 -1\Big) {M^2\over M_\Delta^2}   \ , \\ [3pt]
 \lambda_3(Z) &=&\lambda_3(-\case{1}{2})  +
                {h_{\rm A}^2 \over 9} 
                \Big( Z^2 -Z -\case{3}{4}\Big) 
                   {M^2\over M_\Delta^2}      \ . \label{eq:zlambda3}  
\end{eqnarray}
We shall quote parameters obtained with $Z=-{1\over 2}$
and the parameters for other values of 
$Z$ can be obtained from Eqs.~(\ref{eq:zbeta}) 
to (\ref{eq:zlambda3}). We have verified this numerically.

We use the standard labelling for isospin-spin partial wave channels,
namely $\alpha\equiv(l,2I,2J)$ where $l$ is the orbital angular momentum,
$I$ is the total isospin,
and $J=l\pm{1\over 2}$ is the total angular momentum.
The elastic scattering amplitude 
\begin{equation}
      f_\alpha = {1\over |\bbox{q}|}e^{i\delta_\alpha}
                 \sin \delta_\alpha        \label{eq:f}
\end{equation}
is obtained from the amplitudes $A^\pm$ and $B^\pm$  by 
the usual partial wave expansion\cite{GASIO66}.
Here $\delta_\alpha$ is the phase shift of the $\alpha$ partial wave.

Unitarity requires $f_\alpha$ to take the complex structure in 
Eq.~(\ref{eq:f}). However, $f_\alpha$ is real
in a tree-level approximation to the scattering amplitude.
We may recover unitarity by obtaining the phase shifts 
from two common methods.
The first assumes that  the calculated  $f_\alpha$  is simply the real 
part of Eq.~(\ref{eq:f}). The second 
introduces a $K$ matrix given by \cite{EW88}
\begin{equation}
      f_\alpha = {K_\alpha \over 1 - i |\bbox{q}| K_\alpha}
\quad{\rm where}\quad     
K_\alpha = {1\over |\bbox{q}|} \tan\delta_\alpha \ .     \label{eq:Kf}
\end{equation}
The calculated real tree-level amplitude $f_\alpha$
is then assumed to actually be $K_\alpha$, which is true 
for $|\bbox{q}|$ small enough. 
For sufficiently small phase shifts, the two
methods yield the same answer because 
$\sin \delta_\alpha \approx \tan \delta_\alpha \approx \delta_\alpha$.
However, near the resonance region where $\delta_\alpha\sim \pi/2$,
the $K$-matrix method is preferred for the following simple 
reason. (We note that Goudsmit {\it et al.}\cite{GOUDSMIT94} 
have proposed a justification for the $K$-matrix method.)

First, for energies near a resonance, 
the amplitude in the resonant channel
takes the relativistic Breit-Wigner form. Taking
the $P33$ channel as an example, we have\cite{EW88}
\begin{equation}
    |\bbox{q}|  f_{P33}^{\rm BW} = {M_\Delta \Gamma_\Delta
             \over M_\Delta^2 -s - i M_\Delta \Gamma_\Delta}
                     \ ,    \label{eq:BW}
\end{equation}
where $\Gamma_\Delta$ is the $\Delta$ width. Eqs.~(\ref{eq:f})
and (\ref{eq:BW}) lead to
\begin{equation}
  \tan \delta_{P33} =   {M_\Delta \Gamma_\Delta
             \over M_\Delta^2 -s  }
                     \ . \label{eq:Tan}
\end{equation}
Next, we expect that the tree-level amplitude can be obtained
by setting the
imaginary part of the denominator of  Eq.~(\ref{eq:BW}) to zero:
\begin{equation}
    |\bbox{q}|  f_{P33}^{\rm tree} = {M_\Delta \Gamma_\Delta
             \over M_\Delta^2 -s }
                     \ ,    \label{eq:tree}
\end{equation}
and this is indeed obtained by retaining only the pole contribution 
of Eqs.~(\ref{AplusD}) to
(\ref{BminusD}) and using the partial wave expansion.
Finally, given the tree amplitude Eq.~(\ref{eq:tree}),
the correct phase shift of Eq.~(\ref{eq:Tan})
is obtained  by the $K$-matrix method. Thus,
while the two methods do not differ for small phase shifts
in the nonresonant channels, the $K$-matrix method
is also good on resonance.
We therefore use the $K$-matrix method here.

\begin{figure}
 \setlength{\epsfxsize}{6.0in}
  \centerline{\epsffile{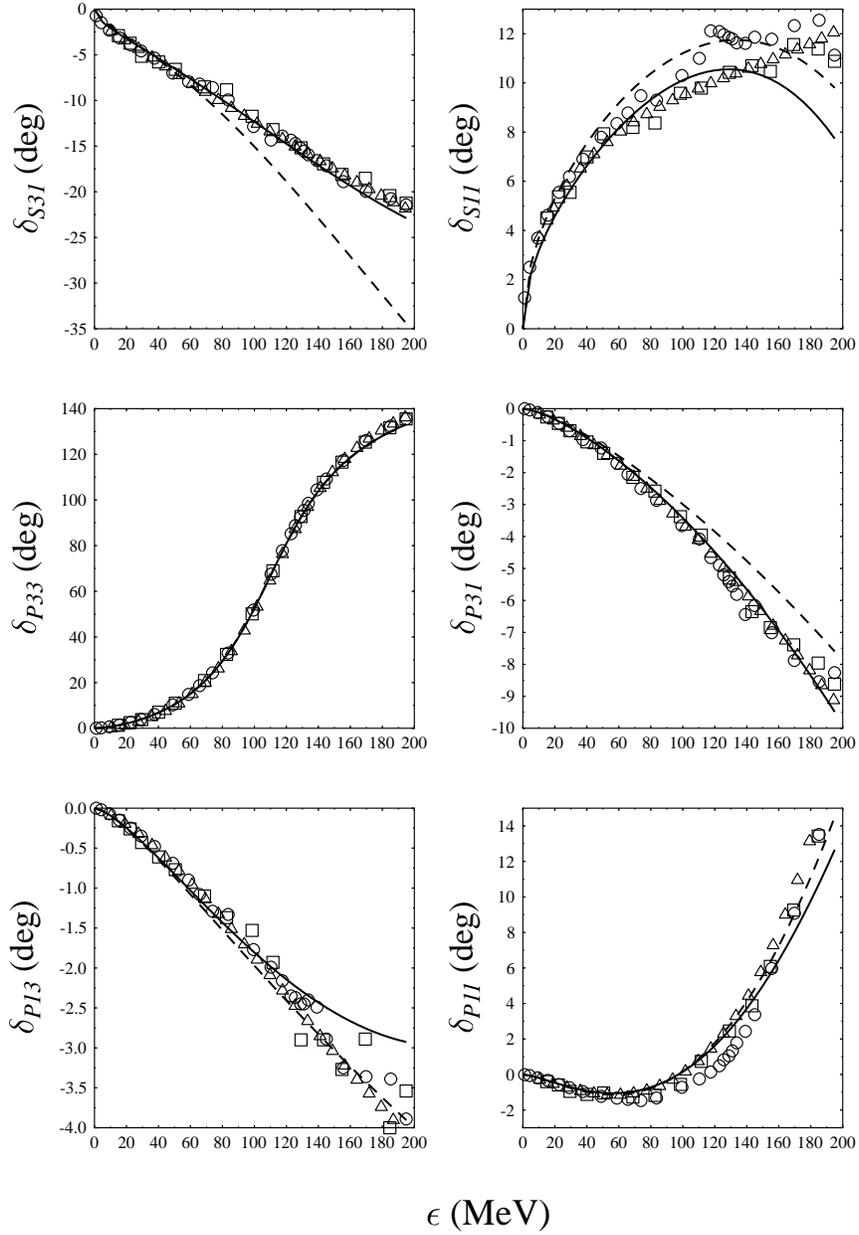}}
\vspace{-1.2in}
\caption{
  The calculated $S$- and $P$-wave phase shifts as functions of the pion c.m.
kinetic energy. The phase-shift data from 
Arndt\protect\cite{ARNDT}
(triangles), Bugg\protect\cite{BUGG} (squares), and Koch and
Pietarinen\protect\cite{KP80} (circles) are also shown.}
 \label{fig:two}
\end{figure}

In our calculations we choose the standard values 
$M=939\,$MeV, $M_\Delta=1232\,$MeV, and $m_\pi=139\,$MeV.
We also take\cite{PDT} $f_\pi = 92.4\,$MeV from charged pion decay, 
$g_{\rm A} = 1.26$ from
neutron $\beta$ decay, and $h_{\rm A}=1.46$ from the $\Delta$
width, $\Gamma_\Delta = 120 \,$MeV; allowing $g_A$ and $h_A$
to vary does not improve the fit. 
We first consider an $O(Q^2)$ approximation to the $T$ matrix
which neglects ${\cal L}_4$.
The four parameters listed in Table 1 were obtained by a $\chi^2$ fit to
the data of Arndt \cite{ARNDT} for
pion c.m.  kinetic energies between $10$ and $150$ MeV.
Because negligible error bars are given in the data at low energies, 
we assign all the data points the same relative weight.
In Fig.~\ref{fig:two}, we plot the calculated $S$- and $P$-wave phase 
shifts (dashed curves), along with the data to which we fit, 
as a function of the pion c.m.  kinetic energy; we also display older data 
from Bugg\cite{BUGG} and from Koch and Pietarinen\cite{KP80}.
The calculation is in good agreement with
the data up to 50 MeV, but beyond this energy the fit deteriorates for 
three of the partial waves.
The value of $\chi^2$ is unity for a relative weight of 15\%
which is a measure of the accuracy of the fit.
The threshold (vanishing pion kinetic energy)
$S$-wave scattering lengths ($a_{2I}$)
and the $P$-wave scattering volumes ($a_{2I\,2J}$)
are given in Table~\ref{tab:two}. The difference between the data from 
Refs. \cite{ARNDT} and \cite{KP80} gives an indication of the error
in the absence of a more reliable estimate. 
As regards theoretical predictions, apart from $a_{13}$ which is closer 
to the older value \cite{KP80},
the $O(Q^2)$ results agree nicely with Ref. \cite{ARNDT} which is to
be expected since they are the zero energy extrapolation of the data we 
have fitted.

We now include ${\cal L}_4$, which involves 
five additional parameters ($\lambda_1$ to $\lambda_5$), to take 
the tree approximation to $O(Q^3)$.
The results are indicated by the solid curve in Fig. 2 which gives a good fit
(with a relative weight of 8\% for $\chi^2 =1$) out to 150 MeV. In fact
only the $S_{11}$ and $P_{13}$ phase shifts deviate significantly from
the data in the 150 --200 MeV range. Of course the rather precise 
agreement for $\delta_{P33}$ is strongly influenced by the phenomenological 
$K$-matrix unitarization. This forces the phase shift to be $\pi/2$ at 
$s=M_{\Delta}^2$ corresponding to a c.m. energy of 127 MeV.
As regards the threshold results given in Table II, the predictions are a 
little closer to the data than at $O(Q^2)$ with the exception of $a_1$.
In this connection it is instructive to examine the isoscalar and isovector 
$S$ wave scattering lengths, $(b_0,\ b_1)$.  A recent 
determination \cite{SIGG96} gave $(-0.008\pm0.007,\ -0.096\pm0.007)$
in units of $m_\pi^{-1}$,
in substantial agreement with Refs. \cite{ARNDT,KP80}; note that Arndt 
favors a value of $b_0$ consistent with zero. At $O(Q^2)$ we obtain
$(0.007,\ -0.081)$ and at $O(Q^3)$ $(-0.010,\ -0.077)$.
Thus the isoscalar $b_0$, which is zero in the chiral limit, 
has improved by going to $O(Q^3)$, while the magnitude of $b_1$ remains too
small.

\begin{table}[tbh]
\caption{Parameters from  fits to the $S$- and $P$-wave phase shifts.
  }
\vspace{.1in}
\begin{tabular}[t]{cccccccccc}
 fit  & $\beta_\pi$  & $\kappa_\pi$ & $\kappa_1  $ & $\kappa_2 $  
      & $\lambda_1 $ & $\lambda_2 $ & $\lambda_3 $ & $\lambda_4 $ 
      & $\lambda_5 $    \\
                        \hline
$O(Q^2)$ & $-$0.1960  &    0.5001 &    0.3061 & $-$0.9328 
         &            &           &           &         
         &              \\
$O(Q^3)$ & $-$0.1376  &    0.5301 &    0.7431 & $-$0.5799 
         &    0.3650  & $-$0.3239 & $-$0.0401 &    0.6334
         &  $-$0.4347
\end{tabular}
\label{tab:one}
\end{table}

Apart from $\lambda_3$ which has little influence on the fit,
the $O(Q^3)$ parameters listed in Table~\ref{tab:one} are
of order unity although Eqs. (\ref{eq:zbeta}) 
to (\ref{eq:zlambda3}) show that, while the fit is independent
of $Z$, the actual parameter values will depend on $Z$.
The pseudoscalar coupling with parameter
$\lambda_1$ allows the effective $\pi NN$ coupling constant 
to be adjusted in the $O(Q^3)$ fit. From the Goldberger-Treiman 
relation, our values for $g_{\rm A}$ anf $f_\pi$ correspond to
a $\pi NN$ coupling, $g_{\pi NN}=$12.8 which is a little lower than 
the value of 13.1 obtained by Arndt {\it et al.} \cite{ARNDT94}. When
the $\lambda_1$ term is included $g_{\pi NN}$ decreases slightly to 12.6.
We will not comment on the sigma term since this requires extrapolation
to the unphysical region which may not be reliable with this tree-level
model.

With 9 parameters our $O(Q^3)$ calculation deviates from the data 
only beyond 150 MeV c.m. energy. At the higher energies we do a little 
better than Goudsmit {\it et al.} \cite{GOUDSMIT94} who have 7 parameters 
and fit to 75 MeV. The calculation of Boffinger and Woolcock
\cite{BOFINGER91}, which is an improved version of Ref. \cite{OLSSON75},
contains 10 parameters and produces a fit which is similar to ours but a 
little better at energies $\sim200$ MeV. The remaining models
\cite{PEARCE91,GROSS93} have a larger number of parameters (14) and 
correspondingly fit to significiantly higher energies. 

In conclusion, we have discussed a chiral lagrangian involving just 
the basic $N$, 
$\pi$ and $\Delta$ fields, with a series of terms representing a 
momentum expansion. We find that a tree-level calculation with this 
model represents the data as well as other models with a similar number of 
parameters. Further we have confirmed by explicit calculation that
the $Z$ parameter of the $\pi N\Delta$ vertex is irrelevant if a 
sufficiently general lagrangian is employed.
Of course it would be more satisfactory if a unitary scattering amplitude 
emerged naturally, rather than being imposed phenomenologically. Such would 
be the case if loops were calculated in heavy baryon chiral perturbation 
theory and work in this direction is in progress.

We acknowledge support from the Department of Energy under grant 
No. DE-FG02-87ER40328.

\begin{table}[tbh]
\caption{The calculated $S$-wave scattering lengths and  $P$-wave
scattering  volumes for the $O(Q^2)$ and $O(Q^3)$ fits compared 
with the data of Refs.~\protect\cite{ARNDT} and \protect\cite{KP80}.
The scattering lengths and volumes are in units of
$m_\pi^{-1}$ and $m_\pi^{-3}$ respectively.  
  }
\vspace{.1in}
\begin{tabular}[t]{cdddd}
 length/volume & $O(Q^2)$  & $O(Q^3)$ & Ref.~\cite{ARNDT} 
               &Ref.~\cite{KP80}    \\
                        \hline
$a_1$    &    0.169 &    0.144 &    0.175 & 0.173\\
$a_3$    & $-$0.074 & $-$0.087 & $-$0.087 & $-$0.101\\
$a_{11}$ & $-$0.074 & $-$0.071 & $-$0.068 & $-$0.081\\
$a_{13}$ & $-$0.032 & $-$0.031 & $-$0.022 & $-$0.030\\
$a_{31}$ & $-$0.038 & $-$0.040 & $-$0.039 & $-$0.045\\
$a_{33}$ &    0.212 &    0.209 &    0.209 &    0.214
\end{tabular}
\label{tab:two}
\end{table}

\end{document}